\documentclass[11pt,a4paper]{article}
\usepackage{geometry}
\usepackage{makeidx}         
\usepackage{graphicx}        
                             
\usepackage{tikz}
\DeclareRobustCommand\full  {\tikz[baseline=-0.6ex]\draw[thick] (0,0)--(0.5,0);}
\DeclareRobustCommand\dashed{\tikz[baseline=-0.6ex]\draw[thick,dashed] (0,0)--(0.54,0);}
\usepackage{multicol}        
\usepackage{amsmath,amsfonts,amssymb}
\usepackage{bbm}
\usepackage{bm}
\usepackage[ruled,norelsize,vlined]{algorithm2e}
\usepackage[colorlinks,linkcolor=black,citecolor=blue,urlcolor=blue]{hyperref}
\makeatletter
\newcommand*{\rom}[1]{\expandafter\@slowromancap\romannumeral #1@}
\makeatother

\makeindex             

\newcommand{\R}{ {\mathbb{R}} }

\newcommand{\bx}{ {\bf x} }
\newcommand{\bP}{ {\bf P} }

\newcommand{\bs}{ {\bf s} }
\newcommand{\bU}{ {\bf U} }
\newcommand{\by}{ {\bf y} }
\newcommand{\bX}{ {\bf X} }

\newcommand{\bD}{ {\bf D} }
\newcommand{\bI}{ {\bf I} }

\newcommand{\bV}{ {\bf V} }

\newcommand{\bG}{ {\bf G} }
\newcommand{\bW}{ {\bf W} }

\newcommand{\ba}{ {\bf a} }

\newcommand{\bgamma}{ {\boldsymbol \gamma} }
\newcommand{\bepsilon}{ {\boldsymbol \epsilon} }
\newcommand{\bvarepsilon}{ {\boldsymbol \varepsilon} }

\newcommand{\bGamma}{ {\boldsymbol \Gamma} }
\newcommand{\bDelta}{ {\boldsymbol \Delta} }

\newcommand{\bmu}{ {\boldsymbol \mu} }

\newcommand{\bOmega}{ {\boldsymbol \Omega} }
\newcommand{\bomega}{ {\boldsymbol \omega} }

\newcommand{\bxi}{ {\boldsymbol \xi} }
\newcommand{\bz}{ {\bf z} }
\newcommand{\btheta}{ {\boldsymbol \theta} }

\newtheorem{theorem}{Theorem}
\newtheorem{corollary}{Corollary}

\begin{document}

\title{Variational inference for the smoothing distribution in dynamic probit models}
\date{}
\author{Augusto Fasano\footnote{Collegio Carlo Alberto and ESOMAS Department, C.so Unione Sovietica 218/bis, Turin.\newline e-mail: augusto.fasano@unito.it} \and Giovanni Rebaudo \footnote{Department of Statistics and Data Sciences, the University of Texas at Austin, TX 78712-1823.\newline e-mail: giovanni.rebaudo@austin.utexas.edu}}

\maketitle
\vspace*{-1cm}
\begin{abstract}
	Recently, \cite{fasano2019closed} provided
	closed-form expressions for the filtering, predictive and smoothing distributions of multivariate dynamic probit models, leveraging on unified skew-normal
	distribution properties.
	This allows to develop algorithms to draw independent and identically distributed
	samples from such distributions, as well as sequential Monte Carlo procedures for the filtering and predictive distributions, allowing to overcome computational bottlenecks that may arise for large sample sizes.
	In this paper, we briefly review the above-mentioned closed-form expressions, mainly focusing on the smoothing distribution of the univariate dynamic probit. 
	We develop a variational Bayes approach, extending the partially factorized mean-field variational approximation introduced by \cite{fasano2019scalable} for the static binary probit model to the dynamic setting.
	Results are shown for a financial application.\\
	
	\noindent	\textit{Some key words:}
	Dynamic Probit Model, Hidden Markov Model, Variational Inference, Unified Skew-Normal Distribution
\end{abstract}

\section{Introduction}
\label{sec:1}
Let us consider a hidden Markov model with binary observations $y_{t}\in \{0;1\}^m$, $t=1,\ldots,n$, and state variables $\btheta_{t}=(\theta_{1t}, \ldots, \theta_{pt})^{\intercal} \in \R^p$. 
Adapting the notation proposed in, e.g., \cite{petris2009} to our setting, we aim to develop a novel variational approximation for the joint smoothing distribution in the following dynamic probit model
\begin{eqnarray}
&&p(y_{t} \mid \btheta_{t})= \Phi((2y_t-1)\bx_t^\intercal\btheta_{t}), \label{eq1}\\ 
&&\btheta_t=\bG_{t}\btheta_{t-1}+\bvarepsilon_t, \quad \bvarepsilon_t \sim  \mbox{N}_p({\bf 0}, \bW_t), \quad t=1 \ldots, n, \label{eq2}
\end{eqnarray}
with $\btheta_0 \sim \mbox{N}_p(\ba_0, \bP_0)$, $\{\bvarepsilon_t \}_{t\ge 1} \perp \{\btheta_t \}_{t\ge 0}$ and $\bepsilon_{t_1}\perp \bvarepsilon_{t_2}$ for any $t_1\ne t_2$.
In \eqref{eq1}, $ \Phi(\cdot)$ is the cumulative distribution function of the standard normal distribution, while $\bx_t$ represents a known covariate vector. 
In the following, we set $\ba_0=\boldsymbol{0}$ to ease notation.

Representation \eqref{eq1}--\eqref{eq2} can be alternatively obtained via the dichotomization of an underlying state-space model for the univariate Gaussian time series $z_t\in \R$,  $t=1,\ldots, n$, which is regarded, in  econometric applications, as a set of time-varying utilities.
Indeed, adapting classical results from static probit regression \cite{Albert_1993}, model  \eqref{eq1}--\eqref{eq2} is equivalent to
\begin{eqnarray}
y_{t}&&={\mathbbm{1}}(z_t > 0)\label{eq3}\\
z_t &&= \bx_t^\intercal\btheta_{t} + \eta_t, \quad \eta_t\sim \mbox{N}(0,1),\label{eq4}\\ 
\btheta_t&&=\bG_{t}\btheta_{t-1}+\bvarepsilon_t, \quad \bvarepsilon_t \sim \mbox{N}_p({\bf 0}, \bW_t), \ t=1 \ldots, n, \label{eq5}
\end{eqnarray}
having $\btheta_0 \sim \mbox{N}_p(\boldsymbol{0}, \bP_0)$, $\{\eta_t \}_{t\ge 1}\perp\{\bvarepsilon_t \}_{t\ge 1}$ and $\eta_{t_1}\perp \eta_{t_2}$ for any $t_1\ne t_2$.

As is clear from model  \eqref{eq4}--\eqref{eq5}, if $\bz_{1:n}=(z_1, \ldots, z_t)^{\intercal}$ were observed, then, calling $\btheta_{1:n}=(\btheta_1^\intercal,\ldots,\btheta_n^\intercal)^\intercal$, 
the joint smoothing density $p(\btheta_{1:n} \mid \bz_{1:n})$ and its marginals $p(\btheta_{t} \mid \bz_{1:n})$, $t \leq n$, could be obtained in closed-form by Gaussian-Gaussian conjugacy \cite{petris2009}.
However, in \eqref{eq3}--\eqref{eq5} only a dichotomized version $y_t$ of $z_t$ is available.
Thus the smoothing density is $p(\btheta_{1:n} \mid \by_{1:n})$, which is not Gaussian.

\section{Literature review}
\label{sec:2}
In the context of static probit regression, \cite{Durante2018} recently proved that the posterior distribution for the probit coefficients, under either Gaussian or unified skew-normal (\textsc{sun}) \cite{arellano_2006} priors, is itself a \textsc{sun} with parameters that can be derived in closed-form.
Leveraging these findings, \cite{fasano2019closed} showed that also in the more challenging multivariate dynamic probit setting, the filtering, predictive and smoothing densities of the state variables have \textsc{sun} kernels. 
We recall that a random vector $\btheta \in \R^q$ has \textsc{sun} distribution, $\btheta \sim \mbox{\textsc{sun}}_{q,h}(\bxi,\bOmega,\bDelta,\bgamma,\bGamma)$, if its density function $p(\btheta)$ can be expressed as
\vspace{-7pt}

\begin{equation*}
\phi_q(\btheta -\bxi; \bOmega) \frac{\Phi_h[\bgamma+\bDelta^\intercal \bar{\bOmega}^{-1} \bomega^{-1}(\btheta-\bxi); \bGamma{-}\bDelta^{\intercal}\bar{\bOmega}^{-1}\bDelta ]}{\Phi_h(\bgamma;\bGamma)},
\end{equation*}
where the covariance matrix $\bOmega$ of the Gaussian density $\phi_q(\btheta -\bxi; \bOmega)$ can be decomposed as $\bOmega=\bomega \bar{\bOmega} \bomega$, i.e.\ by rescaling the correlation matrix $\bar{\bOmega}$ via the diagonal scale matrix $ \bomega=(\bOmega \odot {\bf I}_q)^{1/2}$, with $\odot$ denoting the element-wise Hadamard product.
See \cite{arellano_2006} for additional details on the \textsc{sun} distribution.

From now on, $\bOmega$ will actually denote the covariance matrix of the zero-mean normally distributed vector $\btheta_{1:n}$.
Even though this might seem an abuse of notation with respect to the \textsc{sun} density above, we show that this matrix actually coincides with the second parameter of the \textsc{sun} joint smoothing density reported in Theorem \ref{thm:JointSmoothing} below.
By the recursive formulation \eqref{eq2}, we have that $\btheta_{1:n}$ is normally distributed thanks to closure properties of Gaussian random variables with respect to linear transformations, while $\bOmega$ shows the following block structure.
Calling $\bG_{l}^{t}=\bG_t \cdots \bG_l$, $l\le t-1$, $\bOmega$ is formed by $(p \times p)$-dimensional blocks $\bOmega_{[tt]}=\mbox{var}(\btheta_t)=\bG^{t}_1 \bP_0 \bG^{t\intercal}_1+\sum_{l=2}^t\bG^{t}_l \bW_{l-1}\bG^{t\intercal}_l+\bW_t$, for $t=1, \ldots,n$, and  $\bOmega_{[tl]}=\bOmega^{\intercal}_{[lt]}=\mbox{cov}(\btheta_t, \btheta_l)=\bG_{l+1}^{t}\bOmega_{[ll]}$, for $t>l$.
As a direct consequence of Theorem 2 in \cite{fasano2019closed} adapted to the simpler model \eqref{eq1}-\eqref{eq2}, the following theorem holds.
\begin{theorem}
\label{thm:JointSmoothing}
	Under model \eqref{eq1}--\eqref{eq2}, the joint smoothing distribution has the form
	\begin{eqnarray*}
	(\btheta_{1:n} \mid \by_{1:n}) 
	\sim \mbox{\textsc{sun}}_{p \cdot n, n}(\boldsymbol{0},\bOmega_{1:n\mid n}, \bDelta_{1:n\mid n},  \boldsymbol{0}, \bGamma_{1:n\mid n}), \label{eq:JointSmoothing}
	\end{eqnarray*}
	with $\bOmega_{1:n\mid n}=\bOmega$, $\bDelta_{1:n\mid n}=\bar{\bOmega}\bomega\bD^{\intercal}\bs^{-1}$, $\bGamma_{1:n\mid n}=\bs^{-1}(\bD\bOmega\bD^{\intercal}+\bI_n)\bs^{-1}$,
	where $\bD$ is an $n\times (p\cdot n)$ block-diagonal matrix having block entries $\bD_{[tt]} = (2y_t-1)\bx_t^\intercal$, $t=1,\ldots,n$,  $\bs=[(\bD\bOmega\bD^{\intercal}+\bI_n) \odot \mbox{\bf I}_{n}]^{1/2}$ and $\bI_n$ defines the $n$-dimensional identity matrix.
\end{theorem}
By Theorem \ref{thm:JointSmoothing} and the additive representation of the \textsc{sun} \cite{arellano_2006}, we can get the following probabilistic characterization, which can be used to draw i.i.d.\ samples from the smoothing distribution as in Algorithm \ref{algo1}:
\begin{equation*}
		(\btheta_{1:n} \mid \by_{1:n}) \stackrel{\text{d}}{=} \bomega_{1:n\mid n}(\bU_{0 \ 1:n\mid n}+\bDelta_{1:n\mid n} \bGamma_{1:n\mid n}^{-1} \bU_{1 \ 1:n\mid n}),
\end{equation*}
with $\bU_{0 \ 1:n\mid n}\sim \mbox{N}_{p\cdot n}({\bf 0},\bar{\bOmega}_{1:n\mid n}- \bDelta_{1:n\mid n}\bGamma_{1:n\mid n}^{-1}\bDelta_{1:n\mid n}^{\intercal})$, while $\bU_{1 \ 1:n\mid n}$ 
is distributed according to a
 $\mbox{N}_{n}({\bf 0},\bGamma_{1:n\mid n})$ truncated below $\boldsymbol{0}$.
From this representation, we see that the most computationally demanding part of drawing i.i.d.\ samples is sampling from an $n$-variate truncated Gaussian (point [\rom{2}] of Algorithm 1). Although recent results \cite{botev_2017} allow efficient simulation in small-to-moderate time series, this i.i.d.\ sampler
might become computationally impractical for longer time series.
In this paper, we propose a variational approximation for the smoothing distribution to overcome such computational issues.
This approximation is based on methods developed by \cite{fasano2019scalable}, which we extend here to the dynamic setting.

\setlength{\textfloatsep}{10pt}
\begin{algorithm*}[b]
	\caption{\makebox[\linewidth][l]{Independent and identically distributed sampling from $p(\btheta_{1:n} \mid \by_{1:n})$}}
	\label{algo1}
	\vspace{3pt}
	\begin{description}
		\footnotesize
		\vspace*{-2pt}\item{{\bf [\rom{1}]}  Sample $\bU^{(1)}_{0 \ 1:n\mid n}, \ldots, \bU^{(R)}_{0 \ 1:n\mid n}$ independently from  a $\mbox{N}_{p\cdot n}({\bf 0},\bar{\bOmega}_{1:n\mid n}- \bDelta_{1:n\mid n}\bGamma_{1:n\mid n}^{-1}\bDelta_{1:n\mid n}^{\intercal})$.}
		\vspace{-5pt}\item{{\bf [\rom{2}]} Sample $\bU^{(1)}_{1 \ 1:n\mid n}, \ldots, \bU^{(R)}_{1 \ 1:n\mid n}$ independently from  a $\mbox{N}_{n}({\bf 0},\bGamma_{1:n\mid n})$, truncated below $\boldsymbol{0}$.} 
		\vspace{-5pt}\item{{\bf [\rom{3}]} Compute $\btheta^{(1)}_{1:n\mid n}, \ldots, \btheta^{(R)}_{1:n\mid n}$ via $\btheta^{(r)}_{1:n\mid n}=\bomega_{1:n\mid n}(\bU^{(r)}_{0 \ 1:n\mid n}+\bDelta_{1:n\mid n} \bGamma_{1:n\mid n}^{-1} \bU^{(r)}_{1 \ 1:n\mid n})$ for each  $r$.}
		\vspace*{-2pt}
	\end{description}
	\vspace{-5pt}
\end{algorithm*}

\section{Variational approximation for the smoothing distribution}
\label{sec:3}
\cite{fasano2019scalable} recently introduced a partially factorized mean-field variational Bayes (\textsc{pfm-vb}) approximation for static probit models, which allows to perform approximate posterior inference without incurring in computational issues arising from the i.i.d.\ sampling. See \cite{Durante2018} for details.
Such a procedure has also been extended to categorical observations \cite{fasano2020}, providing notable approximation accuracy, especially in high dimensional settings.
In this section, we adapt such results to develop a variational procedure for approximate inference on the smoothing distribution in dynamic probit models.
Adapting \cite{fasano2019scalable}, our \textsc{pfm-vb} procedure aims at providing a tractable approximation for the joint posterior density $p(\btheta_{1:n},\bz_{1:n} \mid \by_{1:n})$ of the states vector $\btheta_{1:n}$ and the partially observed variables $\bz_{1:n}=(z_1, \ldots, z_n)^{\intercal}$, within the \textsc{pfm} class of partially factorized densities $\mathcal{Q}_{\textsc{pfm}}=\{ q_{\textsc{pfm}}(\btheta_{1:n},\bz_{1:n}): q_{\textsc{pfm}}(\btheta_{1:n},\bz_{1:n})=q_{\textsc{pfm}}(\btheta_{1:n}\mid\bz_{1:n})\prod_{i=1}^nq_{\textsc{pfm}}(z_i) \}$.
Differently from classic mean-field (\textsc{mf}) approximations, this enlarged class does not assume independence among $\btheta_{1:n}$ and $\bz_{1:n}$, thus providing a more flexible family of approximating densities.
This form of factorization allows to remove the main computationally demanding part of $p(\btheta_{1:n},\bz_{1:n} \mid \by_{1:n})$, while retaining part of its dependence structure. 
Indeed, adapting \cite{holmes_2006} and letting $\bV=(\bOmega^{-1}+\bX^{\intercal}\bX)^{-1}$ and $\bX$ a $n\times (p\cdot n)$ block-diagonal matrix with block-diagonal entries $\bX_{[tt]} = \bx_t^\intercal$, $t=1,\ldots,n$, the joint density $p(\btheta_{1:n},\bz_{1:n} \mid \by_{1:n})$ under the augmented model \eqref{eq3}-\eqref{eq5} can be factorized as $p(\btheta_{1:n},\bz_{1:n} \mid \by_{1:n})=p(\btheta_{1:n} \mid \bz_{1:n})p(\bz_{1:n} \mid \by_{1:n})$, where $p(\btheta_{1:n} \mid \bz_{1:n})=\phi_{p\cdot n}(\btheta_{1:n}-\bV\bX^\intercal\bz_{1:n}; \bV)$ and $p(\bz_{1:n} \mid \by_{1:n})\propto\phi_n(\bz_{1:n}; \bI_n+\bX\bOmega\bX^{\intercal})\prod_{i=1}^n \mathbbm{1}[(2y_i-1)z_i>0]$ denote the densities of a $p\cdot n$-variate Gaussian and an $n$-variate truncated normal, respectively.
From this, we can note that the main source of intractability comes from the  truncated normal density.

The optimal \textsc{pfm-vb} solution  $q_{\textsc{pfm}}^*(\btheta_{1:n}\mid\bz_{1:n})\prod_{i=1}^nq_{\textsc{pfm}}^*(z_i)$
within $\mathcal{Q}_{\textsc{pfm}}$ is the minimizer of the  Kullback--Leibler (\textsc{kl}) divergence  \cite{Kullback_1951}
\begin{eqnarray*}
{\normalfont  \textsc{kl}}[q_{\textsc{pfm}}(\btheta_{1:n},\bz_{1:n}) \mid \mid p(\btheta_{1:n},\bz_{1:n} \mid \by_{1:n})]&&=\mathbb{E}_{q_{\textsc{pfm}}(\btheta_{1:n},\bz_{1:n})}[\log q_{\textsc{pfm}}(\btheta_{1:n},\bz_{1:n})]\nonumber\\
&&\ \ -\mathbb{E}_{q_{\textsc{pfm}}(\btheta_{1:n},\bz_{1:n})}[\log p(\btheta_{1:n},\bz_{1:n} \mid \by_{1:n})].\quad \ \ 
\label{eq8}
\end{eqnarray*}
Alternatively, it is possible to obtain $q_{\textsc{pfm}}^*(\btheta_{1:n}\mid\bz_{1:n})\prod_{i=1}^nq_{\textsc{pfm}}^*(z_i)  $ by maximizing the evidence lower bound ${\normalfont  \textsc{elbo}}[q_{\textsc{pfm}}^*(\btheta_{1:n}\mid\bz_{1:n})]$. 
See \cite{blei_2017} and \cite{fasano2019scalable} for details.
Adapting Theorem 2 in \cite{fasano2019scalable}, 
it is immediate to obtain the following theorem.

\begin{theorem}
	Under model \eqref{eq1}-\eqref{eq2}, the $\textsc{kl}$ divergence between \mbox{$q_{\textsc{pfm}}(\btheta_{1:n},\bz_{1:n}) \in \mathcal{Q}_{\textsc{pfm}}$} and $p(\btheta_{1:n},\bz_{1:n} \mid \by)$ is minimized at $q^*_{\textsc{pfm}}(\btheta_{1:n}\mid\bz_{1:n}) \prod_{i=1}^n q^*_{\textsc{pfm}}(z_i)$ with
	\begin{eqnarray}
	\begin{split}
	&q^*_{\textsc{pfm}}(\btheta_{1:n} \mid\bz_{1:n})=p(\btheta_{1:n} \mid \bz_{1:n})=\phi_{p\cdot n}(\btheta_{1:n}-\bV\bX^\intercal\bz_{1:n}; \bV), \\
	&q_{\textsc{pfm}}^*(z_i)= \frac{\phi(z_i-\mu^*_i; \sigma^{*2}_i)}{\Phi[(2y_i-1)\mu^*_i/\sigma^{*}_i]}\mathbbm{1}[(2y_i-1)z_i>0], \quad  \sigma^{*2}_i=(1-\bX_{[i,]}\bV \bX_{[i,]}^{\intercal})^{-1},
	\end{split}
	\label{eq9}
	\end{eqnarray}
	where $\bmu^*=(\mu_1^*, \ldots, \mu_n^*)^{\intercal}$ solves the system  \mbox{$\mu^*_i-\sigma^{*2}_i \bX_{[i,]}\bV\bX^{\intercal}_{[-i,]}\bar{\bz}^*_{-i}=0$}, $i=1, \ldots, n$, with $\bX_{[-i,]}$ denoting the matrix $\bX$ with the $i$th row $\bX_{[i,]}$ removed, while $\bar{\bz}^*_{-i}$ is an $n-1$ vector obtained by removing the $i$th element $\bar{z}^*_i=\mu^*_i+(2y_i-1)\sigma^{*}_i\phi(\mu^*_i/\sigma^{*}_i)\Phi[(2y_i-1)\mu^*_i/\sigma^{*}_i]^{-1}$, $i=1, \ldots, n$, from the vector $\bar{\bz}^*=(\bar{z}^*_1, \ldots, \bar{z}^*_n)^{\intercal}$.
	\label{teo_2}
\end{theorem}

\begin{algorithm*}[b]
	\fontsize{10pt}{10pt}\selectfont
	\caption{\mbox{\textsc{cavi} algorithm for ${q}^*_{\textsc{pfm}}(\btheta_{1:n},\bz_{1:n})={q}^*_{\textsc{pfm}}(\btheta_{1:n}\mid\bz_{1:n}) \prod_{i=1}^n {q}^*_{\textsc{pfm}}(z_i)$}}
	\label{algo2}
	\footnotesize
	{[\bf \rom{1}]} For each $i=1, \ldots, n$, set $\sigma^{*2}_i=(1-\bX_{[i,]}\bV \bX_{[i,]}^{\intercal})^{-1}$ and initialize  $\bar{z}^{(0)}_i\in\R$.\\
	\begingroup
	{{[\bf \rom{2}]}      \For(){$t$ \mbox{from} $1$ until convergence of ${\normalfont  \textsc{elbo}}[q_{\textsc{pfm}}^{(t)}(\btheta_{1:n}, \bz_{1:n})]$}
		{
			\For(){$i$ \mbox{from} $1$ \mbox{to} $n$}
			{ 
				{\bf [\rom{2.1}]} Set $\mu^{(t)}_i=\sigma^{*2}_i \bX_{[i,]}\bV\bX^{\intercal}_{[-i,]}(\bar{z}^{(t)}_1, \ldots, \bar{z}^{(t)}_{i-1}, \bar{z}^{(t-1)}_{i+1}, \ldots,  \bar{z}^{(t-1)}_{n})^{\intercal}$.\\
				{\bf [\rom{2.2}]} Set $\bar{z}^{(t)}_i=\mu^{(t)}_i+(2y_i-1)\sigma^{*}_i\phi(\mu^{(t)}_i/\sigma^{*}_i)\Phi[(2y_i-1)\mu^{(t)}_i/\sigma^{*}_i]^{-1}$.}
		}
			
	}
	
	{\bf Output:} ${q}^*_{\textsc{pfm}}(\btheta_{1:n},\bz_{1:n})$ as in Theorem \ref{teo_2}.
	\endgroup
\end{algorithm*}

Algorithm \ref{algo2} shows how to obtain $q^*_{\textsc{pfm}}(\btheta_{1:n}\mid\bz_{1:n}) \prod_{i=1}^n q^*_{\textsc{pfm}}(z_i)$ via a coordinate ascent variational inference (\textsc{cavi}) algorithm that iteratively optimizes each $\mu_i^*$, keeping the rest fixed \cite{blei_2017}.
In addition to retaining part of the dependence structure of the true posterior, the \textsc{pfm-vb} solution also admits closed-form moments, as shown in Corollary \ref{cor1} below, whose proof can be found in \cite{fasano2019scalable}.
If more complex functionals are desired, they can be easily computed via Monte Carlo integration, since, exploiting \eqref{eq8}, in order to get i.i.d.\ samples from ${q}^*_{\textsc{pfm}}(\btheta_{1:n},\bz_{1:n})$ it is sufficient to draw values from $p\cdot n$-variate Gaussians and univariate truncated normals, avoiding the computational issues of the truncated multivariate normals in Algorithm \ref{algo1}.
\begin{corollary}
	\label{cor1}
	Let $q_{\textsc{pfm}}^*(\btheta_{1:n})=\mathbb{E}_{q_{\textsc{pfm}}^*(\bz_{1:n})}[q_{\textsc{pfm}}^*(\btheta_{1:n}\mid\bz_{1:n})]$, then
	$\mathbb{E}_{q_{\textsc{pfm}}^*(\btheta_{1:n})}(\btheta_{1:n})=\bV \bX^{\intercal}\bar{\bz}^*$
	and $\mbox{\normalfont var}_{q_{\textsc{pfm}}^*(\btheta_{1:n})}(\btheta_{1:n})=\bV+\bV\bX^{\intercal}\mbox{\normalfont diag}[\sigma_1^{*2}-(\bar{z}_1^*-\mu_1^*)\bar{z}_1^*, \ldots, \sigma_n^{*2}-(\bar{z}_n^*-\mu_n^*)\bar{z}_n^*]\bX \bV$,
	where $\bar{z}_i^*$, $\mu_i^*$ and $\sigma_i^{*}$, $i=1, \ldots, n$ are defined as in Theorem \ref{teo_2}.
\end{corollary}
\vspace*{-0.5cm}
\section{Financial application}
\label{sec:4}
\begin{figure}[t]
	\centering
	\includegraphics[width=0.9\linewidth,height=0.3\linewidth, trim={0.2cm 0.2cm 0.2cm 0.2cm},clip]{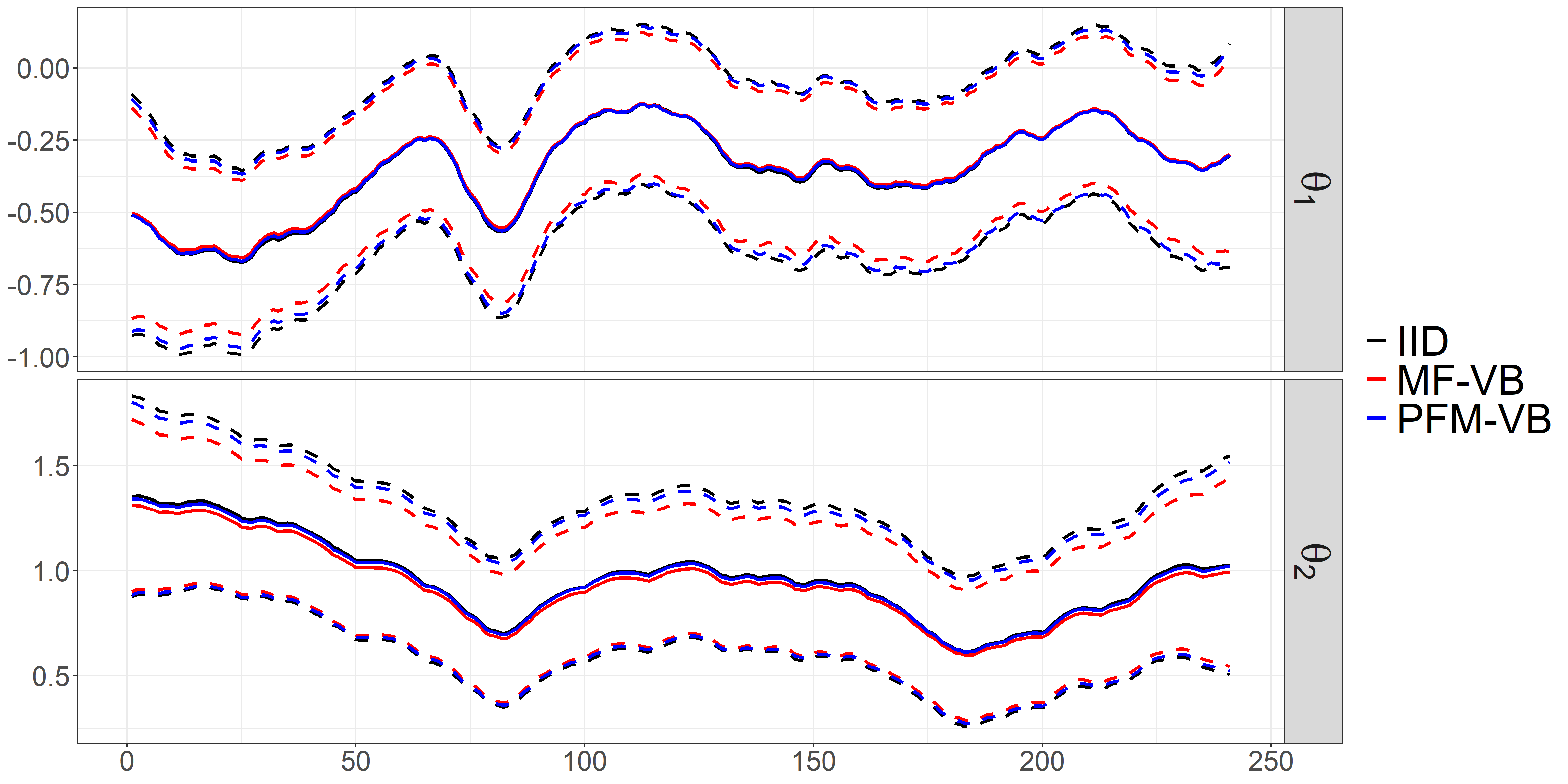}
	%
	%
	\caption{$\mathbb{E}[\btheta_{1:n}\mid\by_{1:n}]$ (\full) and $\mathbb{E}[\btheta_{1:n}\mid\by_{1:n}] \pm \sqrt{\text{var}[\btheta_{1:n}\mid\by_{1:n}]}$ (\dashed) for the i.i.d.\ sampler, the \textsc{mf-vb} algorithm and the \textsc{pfm-vb} solution.}      
	\label{fig:1}
	\vspace*{-0.2cm}
\end{figure}

We illustrate the performance of the variational approximation derived in Section \ref{sec:3} on a financial application considering a dynamic probit regression for the daily opening directions of the  French \textsc{cac}40 stock market index from January 4th, 2018 to December 28th, 2018, for a total of $n=241$ observations.
In this study, $y_t=1$ if the opening value of the \textsc{cac}40 on day $t$ is greater than the corresponding closing value in the previous day, and $y_t=0$ otherwise.
We consider two covariates: the intercept and the opening direction of the \textsc{nikkei}225, regarded as binary covariates $\xi_t$.
Since the Japanese market opens before the French one, $\xi_t$ is available before $y_t$ and, hence, provides a valid predictor for each day $t$.
Thus, with reference to model \eqref{eq1}-\eqref{eq2}, $p=2$ and $\bx_t=(1, \xi_t)^\intercal$. Moreover, we take $\bW_t = \text{diag}(0.01,0.01)$ for every $t$ and $\bP_0=\text{diag}(3,3)$. See \cite{fasano2019closed} for details on the hyperparameters' setting.
The extent of the quality of the \textsc{pfm-vb} approximation is displayed in Figure \ref{fig:1}. There, we plot $\mathbb{E}[\btheta_{1:n}\mid\by_{1:n}]$ and $\mathbb{E}[\btheta_{1:n}\mid\by_{1:n}] \pm \sqrt{\text{var}[\btheta_{1:n}\mid\by_{1:n}]}$, estimated with $10^4$ samples from the i.i.d.\ sampler, with the \textsc{pfm-vb} solution, exploiting Corollary \ref{cor1}, and with a mean-field variational Bayes (\textsc{mf-vb}) approximation, where independence among $\btheta_{1:n}$ and $\bz_{1:n}$ is enforced, by adapting \cite{consonni_2007} to the current setting.
We observe that the \textsc{pfm-vb} approximation---differently from the \textsc{mf-vb}---almost perfectly matches the quantities of interest of the smoothing distribution.
To better understand the improvements of \textsc{pfm-vb} over \textsc{mf-vb}, the average absolute difference in the estimated means of $\theta_{1t}$ and $\theta_{2t}$, $t=1,\ldots,241$, with respect to the ones obtained with the i.i.d.\ sampler are $0.003$ and $0.008$ for the \textsc{pfm-vb} and $0.009$ and $0.031$ for the $\textsc{mf-vb}$, respectively.
Considering the average difference of the log-standard-deviations, we obtain $0.04$ and $0.05$ for the \textsc{pfm-vb}, while these values equal $0.14$ and $0.16$ for the \textsc{mf-vb}, showing a much higher overshrinkage towards $0$.
Finally, the \textsc{pfm-vb} solution allows to compute the desired moments in only 1.1 seconds, similar to \textsc{mf-vb}, showing a much lower computational time than the i.i.d.\ sampler, which requires 115.4 seconds.
Code to produce Figure \ref{fig:1} and additional outputs are available at the following link: \href{https://github.com/augustofasano/Dynamic-Probit-PFMVB}{https://github.com/augustofasano/Dynamic-Probit-PFMVB}.

\begin{paragraph}{Acknowledgments}
	The authors wish to thank Daniele Durante for carefully reading a preliminary version of this manuscript and providing insightful comments.
\end{paragraph}


\end{document}